\documentclass[prb,aps,twocolumn,showpacs,preprintnumbers,amsmath,amssymb,twoside]{revtex4-1}

\usepackage{graphicx}		
\usepackage{pstricks}
\usepackage{epsfig}
\usepackage{dcolumn}		
\usepackage{bm}			
\usepackage{color}
\usepackage{hyperref}
\hypersetup{colorlinks=true,urlcolor=blue,linkcolor=blue,citecolor=blue}

\newcommand{\LCO}{La$_2$CuO$_{4}$}
\newcommand{\LSCO}{La$_{{2-x}}$Sr$_{{x}}$CuO$_{{\rm 4}}$}
\begin{document}
\title{Bimagnon studies in cuprates with Resonant Inelastic X-ray Scattering \\
at the O $K$  edge. II - The doping effect in \LSCO\\}

\author{V. Bisogni$^{1,2}$}
\altaffiliation[Present address: ]{Leibniz Institute for Solid State and Materials Research IFW Dresden, P.O. Box 270116, D-01171 Dresden, Germany}
\author{M. Moretti Sala$^3$}
\altaffiliation[Present address: ]{European Synchrotron Radiation Facility - Bo\^{i}te Postale 220, F-38043 Grenoble, France}
\author{A. Bendounan$^4$}
\altaffiliation[Present address: ]{Synchrotron Soleil, L'Orme des Merisiers, Saint-Aubin - Bo\^{i}te Postale 48, F-91192 Gif-sur-Yvette Cedex, Paris, France}
\author{N.\,B. Brookes$^1$}
\author{G. Ghiringhelli$^3$}
\author{L. Braicovich$^3$}

\affiliation{$^1$ European Synchrotron Radiation Facility - Bo\^{i}te Postale 220, F-38043 Grenoble, France}
\affiliation{$^2$ Leibniz Institute for Solid State and Materials Research IFW
Dresden, Helmholtzstrasse 20, 01069 Dresden, Germany}
\affiliation{$^3$ CNR-SPIN, Dipartimento di Fisica, Politecnico di Milano - Piazza Leonardo da Vinci 32, 20133 Milano, Italy}
\affiliation{$^4$ Paul Scherrer Institute, ETH Zurich and EPF Lausanne - 5232 Villigen PSI, Switzerland}

\date{Received: \today}

\begin{abstract}
We present RIXS data at O $K$ edge from \LSCO~ vs. doping between x=0.10 and  x=0.22 with attention to the magnetic excitations in the Mid-Infrared region. The sampling done by RIXS  is the same as in the undoped cuprates provided the excitation is at the first pre-peak induced by doping. Note that this excitation energy is about 1.5 eV lower than that needed to see bimagnons in the parent compound. This approach allows the study of the upper region of the bimagnon continuum around 450 meV within about one third of the Brillouin Zone around $\Gamma$. The results show the presence of damped bimagnons and of higher even order spin excitations  with almost constant spectral weight at all the dopings explored here. The implications on high T$_c$ studies are briefly addressed.
\end{abstract}

\pacs{78.70.Ck, 78.70.En, 74.72.Gh, 75.30.Ds} 

\maketitle

\section{Introduction}
The study of elementary excitations in cuprate high temperature superconductors is of paramount importance and in this context the spin excitations play a central role. These include the lowest order process i.e. the single propagating magnon, the next order i.e. the bimagnon consisting of the coherent excitation of  two magnons, and all the higher order processes (even and odd). 

This is the second of two papers on bimagnons and even higher order excitations in cuprates studied with Resonant Inelastic X-ray Scattering (RIXS) at the O $K$ edge. In the present paper we investigate the effect of doping by taking  advantage of the results given in the previous paper (referred to as paper I) on \LCO~(LCO) which is prototypical of parent compounds.  The attention will be given to magnetic excitations in the low energy scale. As already done in paper I we stress the importance of these excitations in high T$_c$ superconductors. Whether these excitations are directly involved in the pairing mechanism is still an open problem, but it remains that the high critical temperatures are compatible with the recent RIXS findings on the spin excitation in the whole Brillouin Zone (BZ). \cite{Tacon2011} Although the reading of paper I is greatly useful, the present paper  can also be read  independently provided that some results given in paper I are kept in mind. These are briefly summarized here for convenience of the reader.

Let us consider the bimagnon continuum in an undoped cuprate. The O $K$ edge RIXS gives a sampling of the upper part of this distribution in a region near $\Gamma$ up to roughly 0.4 of the BZ boundary. Both in the (1,0) and in the (1,1) directions of the basal planes there is basically no dispersion with both linear polarizations ($\sigma$ and $\pi$). Also, the intensities are essentially constant at increasing $q$ when $\sigma$ polarization is used. The absence of dispersion or of a very small one is a big difference with respect to the Cu $L_3$ and Cu $K$ edges, where the matrix elements emphasize the lower energy part of the bimagnon continuum, giving a dispersive contribution. The sampling of the upper part of the continuum with O $K$ RIXS is also supported by theory as shown in paper I. This type of sampling at O $K$ emphasizes energy losses peaked typically around 450 meV and makes it possible to work without pushing the experimental resolution to the extreme limits.

In the present paper we demonstrate first that the above rules are valid also in doped samples of the \LSCO~(LSCO) family, provided that the excitation by the incident photons is done at the first pre-edge feature which appears in the absorption spectrum upon doping.\cite{Chen1991,Chen1992,Peets2009} With this provision we study the bimagnon at increasing doping, up to strong overdoping (x = 0.22). The main result is that the spectral weight of the bimagnon and higher order even excitations is basically constant upon doping including strong overdoping, a fact never seen before and easily obtained with the present approach. In the meantime the spectral distribution indicates a damping of the even order excitations at high dopings. This behavior is closely parallel to what seen in Ref.\,\onlinecite{Tacon2011} on single magnon and expands the picture to strong overdoping in the case of bimagnons. Thus it appears that spin excitations are ubiquitous in the form of damped fluctuations; in other words they are present in the whole region of the phase diagram where superconductivity takes place. This correlation will deserve further theoretical investigation possibly stimulated by the present experimental results.

The paper is organized as follows. The experimental information are briefly given in Sec.\,\ref{exp} (more  extended information in paper I), the experimental results are given in Sec.\,\ref{results} and discussed in Sec.\,\ref{discussion}. The conclusions are summarized in Sec.\,\ref{conc}.

\section{Experimental}
\label{exp}
All measurements have been taken at the beamline ID08 of the ESRF with the AXES\cite{axes} spectrometer coupled to the Polifemo\cite{polifemo} monochromator before the entrance slit of the beamline Dragon monochromator. As explained in paper I this is the first of the two setups presented therein and is less luminous so that the measurements have some more noise than in paper I. However the conclusions reported here are not affected by this circumstance. The present data are taken at 110$^\circ$ scattering angle while the undoped LCO spectra of paper I refer to 130$^\circ$. We have found that the parent compound spectra have within the noise the same lineshape at the two scattering angles so that the comparison with the spectra of paper I is significant; nevertheless the comparison with the parent compound presented in this second paper is done for safety with the spectra at 110$^\circ$, taken strictly in the same conditions as in LSCO. All data are taken with a temperature of  20 K. The transferred momentum $q$ is in the direction from (0,0) to (1,0), while going toward grazing incidence.

\begin{figure}
\center{
\resizebox{0.9\columnwidth}{!}{%
 \includegraphics[clip,angle=0]{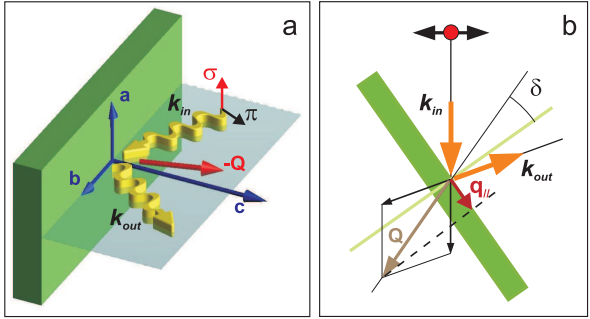}}}
\caption{(Color online). a) Sketch of the RIXS experiment: the sample is in green with its crystallographic axis in blue, while the scattering plane is in light blue. The red and the black arrows indicate the incoming photon polarizations, respectively parallel ($\sigma$) and perpendicular ($\pi$) to the (a,b) plane. In b) the details of the 110$^\circ$ scattering geometry.}
\label{fig1}
\end{figure}

For convenience of the reader we report in Fig.\,\ref{fig1} the geometrical scheme of the experiment. 

The present measurements are taken both on thin films and on bulk crystals. In test experiments, we did not find differences  within the error bars between film and bulk samples at the same doping. The undoped samples, LCO, are the same of paper I. The doped samples of LSCO correspond to x = 0.1, 0.12, 0.145, 0.22 so that we span the range from underdoping to strong overdoping with  intermediate dopings near the ``magic'' 1/8 value and near optimally doping. The samples with x = 0.10 and 0.22 are bulk crystals grown by the traveling floating zone method \cite{Chang2008} and the others are films deposited by laser ablation on SrTiO$_3$. 

\section{Results and discussion}
\subsection{An overview of the experimental results}
\label{results}
\begin{figure}
\center{
\resizebox{0.7\columnwidth}{!}{%
 \includegraphics[clip,angle=0]{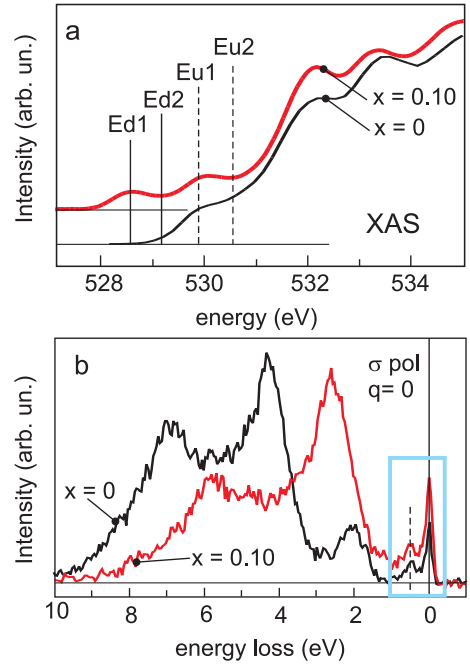}}}
\caption{(Color online). a) O $K$  XAS on x = 0 (black) and x = 0.10 LSCO compounds. The vertical lines mark the excitation energies used for RIXS across the doped peak (solid lines, $Ed1$ and $Ed2$) and the undoped peak (dashed lines, $Eu1$ and $Eu2$). b) RIXS spectra for x = 0 (black) and x = 0.10 (red) LSCO compounds measured respectively by exciting at $Eu1$ and $Ed1$, with $\sigma$ polarization and $q$ = 0. }
\label{fig2}
\end{figure}

The doping has a well known effect on the O $K$ absorption spectrum (XAS) which has consequences on the RIXS measurements. In the XAS, the doping gives rise to a second pre-edge feature at lower energy with respect to the first pre-edge feature of the parent compound. The XAS spectra with x = 0 and x = 0.10 measured with $\sigma$ polarization in the electron yield mode are given in Fig.\,\ref{fig2}(a) in agreement with the literature.\cite{Chen1991,Peets2009} In fact the relative positions of the features are the same; the spectral weights are not coincident because of the different self-absorption/saturation in the electron yield and in the photon yield modes. Thus we took RIXS measurements at the ``doping peak'' ($Ed1$), above this peak ($Ed2$), at the peak typical of the undoping ($Eu1$) and above ($Eu2$).

In the doped samples due to the lower excitation energy the RIXS region is closer to the resonant fluorescence peaks so that the energy window available to RIXS is considerably narrower as shown clearly in Fig.\,\ref{fig2}(b). The region of interest for the present work is still accessible as pointed out by the blue box but a lot of other information is lost upon doping; for example the fluorescence does not allow the observation of the counterpart of the 2 eV peak seen in the parent compound and containing a $dd$-contribution and the excitonic Zhang-Rice singlet.\cite{Okada2002,Ellis2008}

\begin{figure} 
\center{
\resizebox{0.87\columnwidth}{!}{%
 \includegraphics[clip,angle=0]{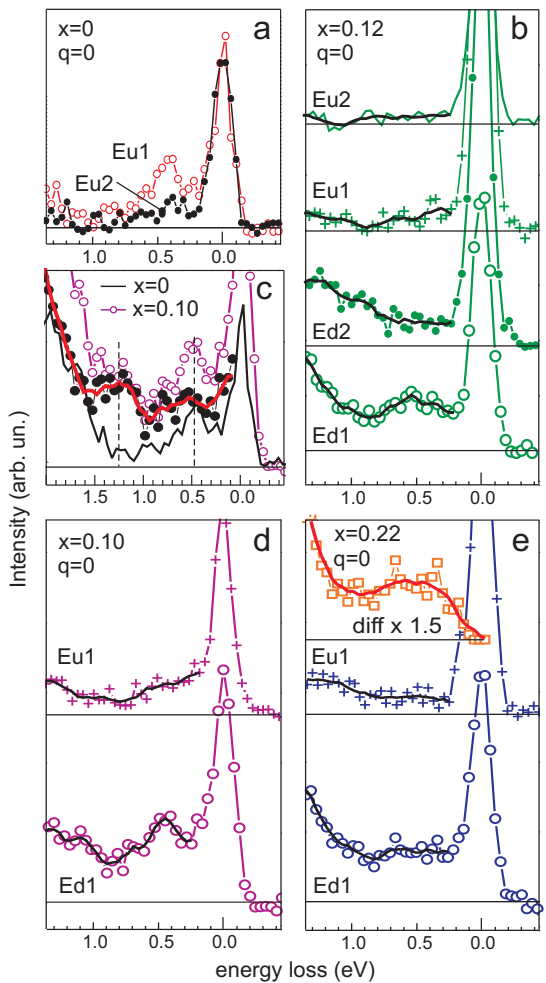}}}
\caption{(Color online). Overview of O $K$  RIXS spectra on LSCO vs. doping, with $\sigma$ polarization and $q$ = 0. a) Spectra on LCO, at $Eu1$ (red) and $Eu2$ (black) excitation energies. b) Stack of spectra on x = 0.12 LSCO, at $Ed1$, $Ed2$, $Eu1$ and $Eu2$ excitation energies. c) Comparison between the x = 0 spectrum (black solid line) at $Eu1$ and the x= 0.10 spectrum (purple open dots) at $Ed1$. Their difference is represented by the black dots. The red curve is a smoothing as a guide to the eye. d) Stack of spectra on x = 0.10 LSCO, at $Ed1$ and $Eu1$. e) Stack of spectra on x = 0.22 LSCO, at $Ed1$ and $Eu1$ (in blue color). The orange square line is the difference between the two spectra, multiplied by 1.5 after subtraction of the elastic line.} 
\label{fig3}
\end{figure}

The effect of the excitation energy is summarized in Fig.\,\ref{fig3}. For comparison we give in Fig.\,\ref{fig3}(a) the results for the undoped case showing as in paper I the resonant excitation in the Mid-Infrared (MI) region at the $Eu1$ energy (in red), in comparison with the $Eu2$ energy. Upon doping, the strongest resonant effect is found by exciting at the new feature $Ed1$. This is particularly evident from Fig.\,\ref{fig3}(b) (x=0.12) where the spectra at the four energies from $Ed1$ to $Eu2$ are given (raw data). In particular the resonance at the undoping energy $Eu1$ is extremely weak at the limit of the noise while the doping resonance at $Ed1$ is very well seen. The resonant spectra at the lowest doping considered here (x = 0.1) and at the x = 0 are compared in Fig.\,\ref{fig3}(c). In both cases the MI feature around 400-450 meV is well seen. The difference spectrum (black dots fitted by the red solid curve) confirms this fact and shows another feature at 1.3 eV. Note that in figure \,\ref{fig3}(c) the normalization is to the same fluorescence contribution. Even by changing normalization the MI and the 1.3 eV features remain present although with different weights. The clear resonance at the pre-edge energy $Ed1$ is a general fact seen at all dopings as evidenced in Fig.\,\ref{fig3}(d) (x = 0.1) and \ref{fig3}(e) (x = 0.22) i.e. including the overdoped regime. In this case the difference spectrum at $Eu1$ and $Ed1$ gives a clear broad resonance shown by the orange curve in Fig.\,\ref{fig3}(e).

\begin{figure}
\center{
\resizebox{0.90\columnwidth}{!}{%
 \includegraphics[clip,angle=0]{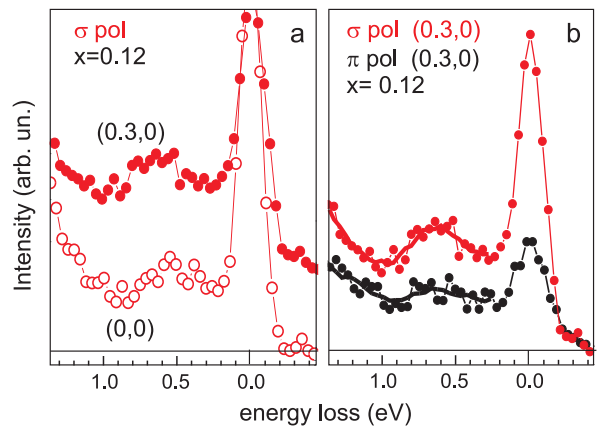}}}
\caption{(Color online). a) Momentum dependence of O $K$  RIXS spectra on x = 0.12 LSCO, for $q$=(0,0) (open dots) and $q$=(0.3,0) (filled dots). b) Polarization dependence at $q$=(0.3,0): red spectrum has been measured with $\sigma$ polarization and black spectrum with $\pi$ polarization.}
\label{fig4}
\end{figure}

The $q$ and the polarization dependence is shown for the case of x = 0.12 in Fig.\,\ref{fig4}(a) where we compare the spectra for the (1,0) direction at $q$ = 0 and $q$ = 0.3 (raw data) which is near the maximum transferred momentum one can obtain in the present setup. The persistence of the MI feature at increasing $q$ is very clear in analogy with the undoped case. Also the polarization dependence (Fig.\,\ref{fig4}(b)) is analogous to the undoped case with a reduction of the MI intensity with $\pi$ polarization at high $q$ even after normalization to the same fluorescence intensity as done in the figure.

\begin{figure}
\center{
\resizebox{0.8\columnwidth}{!}{%
 \includegraphics[clip,angle=0]{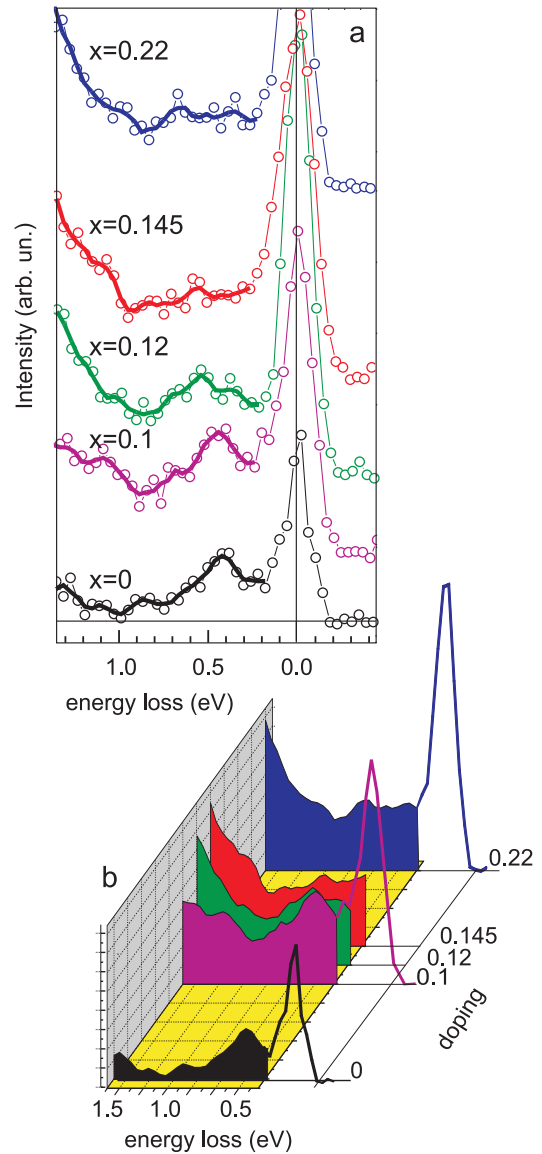}}}
\caption{(Color online). a) Overview of O $K$  RIXS spectra vs. doping, at $q$ = 0 with $\sigma$ polarization. The excitation energy is $Ed1$, except for the x = 0 compound where $Eu1$ has been used. b) Perspective drawing of RIXS spectra on LSCO plotted versus doping. Colored area highlights the inelastic region from 0.25 to 1.5 eV.}  
\label{fig5}
\end{figure}

The doping dependence in the MI region is shown in Fig.\,\ref{fig5} based on the data at $q$ = 0 with $\sigma$ polarization. It is very important to note that this represents the situation also at finite $q$ due to the absence of dispersion. Although most of these spectra are seen also in the previous figures, the direct comparison vs. doping is illuminating. This is done in a stack of spectra and in a perspective drawing comparing different dopings. 

A lot of information on the doping effect can be recovered from this and the preceding figures:

(i) the MI feature centered around 400-450 meV with a tail up to 0.8-1 eV is present at all dopings.

(ii) at x = 0.1 the MI peak has basically the same width as in the undoped parent compound, while a broadening is found at x = 0.14 and above indicating a damping of the MI excitations (see also below). The system at x = 0.12, i.e. near the ``magic'' 1/8 doping, has a feature slightly more extended in energy.

(iii) there is an increase in intensity on going from x = 0 to x = 0.10, i.e. when the excitation energy is changed from $Eu1$ to $Ed1$. 

(iv) a feature at 1.3 eV is seen in the x = 0.1 case while it is hardly seen at higher dopings where it is submerged by the increasing continuum and by the tail of the fluorescence signal.

(v) the intensity of the MI resonance with $Eu1$ energy (i.e. at the undoped XAS feature) is severely reduced at increasing x and is suppressed faster with x in comparison with the decrease of the undoped feature in oxygen $K$ XAS, well documented in the literature.

(vi) at increasing dopings the systems metalize and this is reflected also in the increase of the elastic peak.

\begin{figure}
\center{
\resizebox{0.9\columnwidth}{!}{%
 \includegraphics[clip,angle=0]{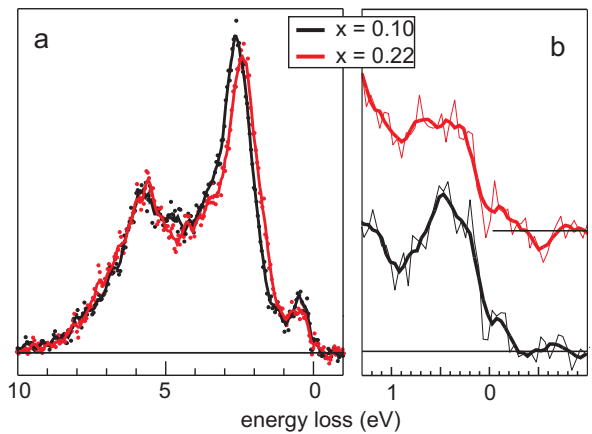}}}
\caption{(Color online). a) Spectra of x=0.1 and x=0.22 LSCO after subtraction of the elastic line. b) An expansion of the region of low energy excitations in the final state; the spectrum with 0.22 doping has an offset for readability.}
\label{fig6}
\end{figure}

Finally we give significant information on the spectra after subtraction of the elastic peak. In particular we compare the minimum (x = 0.10) and maximum (x = 0.22) doping in our measurements. The spectra in the whole energy range including fluorescence are given in Fig.\,\ref{fig6}(a). This demonstrates that the fluorescence does not change with the doping apart a small shift so that the normalization to fluorescence in the previous figures is safe and does not introduce artifacts. The expansion given in Fig.\,\ref{fig6}(b) shows the MI region and adds further information on damping. There is a minor loss of intensity and this happens in the peak region while the total width at the base does not change appreciably so that the width at half height  increases upon doping. This is clear already without any background subtraction and is not influenced by possible presence of  phonon contribution. 

\subsection{Discussion}
\label{discussion}

We concentrate on the MI region which is the main topic of the present paper. The first issue is that the feature seen in this region in doped samples has a strict analogy with that seen in the undoped parent compound. Not only they are almost at the same energy in the final state but they have the same dependence on the transferred momentum and on the incident polarization. Thus the MI feature sampled with O $K$ RIXS is dominated by spin excitations also in the doped compounds. This happens while different incident energies are needed to extract the feature in the two cases. Indeed the spin excitations at Cu $ L _3$ are seen both in undoped and doped systems with the same peak excitation at Cu $L_3$. At O $K$ the situation is different; in LCO the excitation is to unoccupied states which upon doping are considerably higher than the Fermi level in samples which are becoming metallic. Thus the excitation of a doped system at the energy $Eu1$ brings to an intermediate state in which the excited electron has a considerable probability of itinerate away within the lifetime of the core hole. In such a case one obtains a fluorescence photon and not a RIXS contribution. This explains why the $Eu1$ energy is inefficient in giving a RIXS contribution in doped systems so that the excitation at the first pre-edge peak $Ed1$ is needed. In the Cu $ L _3$ case the energy scale is instead much more compressed due to the strong excitonic interaction with the core hole in the intermediate state. Moreover the change of incident energy used in doped systems modifies the cross section for the spin excitation in the final state; it is thus not surprising that we see a jump in intensity in the MI region on going from undoped to doped systems.

\begin{figure}
\center{
\resizebox{0.55\columnwidth}{!}{%
 \includegraphics[clip,angle=0]{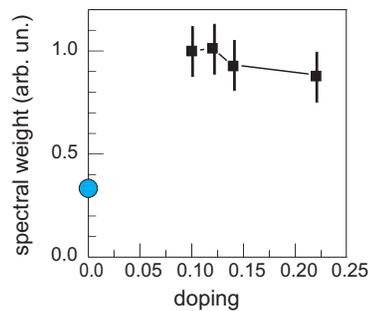}}}
\caption{(Color online). Integral of the difference curves ($Ed1$-$Eu1$) considered between 0.25 and 1 eV  as a function of doping.  The blue dot refers to the undoped case (LCO).}
\label{fig7}
\end{figure}

The doping effect is well represented by the difference between the spectra at the excitation energies $Ed1$ and  $Eu1$; we have already shown as an example the resonating doping contribution at x = 0.22 in Fig.\,\ref{fig3}(e) (in orange). The total contribution defined as the integral of the difference curves ($Ed1$ - $Eu1$) between 0.25 and 1 eV is given vs. doping by the black squares in Fig.\,\ref{fig7}. This shows clearly that the bimagnon is present at all dopings including strong overdoping and its spectral weight is almost constant within the experimental accuracy or more likely is decreasing very slowly with overdoping. As already shown this happens while the bimagnon feature becomes broader and this is the evidence of damping. The conclusion on the basic conservation of spectral weight deals with the upper part of the bimagnon continuum which is the dominant term in O $K$ RIXS. In the parameter space this corresponds to a region with small $q$ and high energy (400-450 meV) which cannot be explored with RIXS at the other edges and which has not been studied with neutrons to the best of our knowledge. In terms of spectral weight the comparison with the parent compound is hardly possible because of the different energy of the incident photons; in fact the blue point representing the undoped material in Fig.\,\ref{fig7} is out of the trend. With Cu $K$ RIXS a somewhat analogous effect on the spectral weight sampled with O $K$ edge (which is a different sampling) has been seen by Ellis $et~al.$  in the underdoped regime,\cite{Ellis2010} leaving the problem open at high doping. Our results give the answer by exploiting the oxygen $K$ edge. 

The present results on bimagnon complement and go along the same direction of the recent observations with Cu $L_3$ RIXS on single paramagnon excitation. Already in Ref.\,\onlinecite{Braicovich2010} on LSCO near x = 1/8 doping we have seen that the phase separation takes place with spin excitations up to about 350 meV in a wide region of the $q$-space. This explains the greater width of the bimagnon peak near 1/8 doping. Moreover in the YBa$_2$Cu$_3$O$_{6+x}$ family damped magnons (i.e. paramagnons) are seen in the great majority of $q$-space, for the explored cases from underdoped regime to slight overdoping. Thus a scenario even more general than in Ref.\,\onlinecite{Tacon2011} is emerging on the ubiquitous presence of spin excitations.\cite{LucioNote} This is important because in Ref.\,\onlinecite{Tacon2011} it has been shown that the spin excitation spectral function extracted from the measurements at Cu $L_3$ in YBa$_2$Cu$_3$O$_{6+x}$ family can justify a very high T$_c$ if used to account for bosonic glue in the electron pairing. Establishing within the same scheme the role of high order spin fluctuations measured here is a fascinating perspective which could be stimulated by the present experimental results. 

Finally we briefly comment for completeness the feature seen at 1.2-1.3 eV upon doping. This is clearly the counterpart of the 1.3 eV feature seen with Cu $K$ RIXS. The present results are fully consistent with the interpretation given in Ref.\,\onlinecite{Ellis2011} where the feature is assigned to a transition from $d$-states to a band intersecting the Fermi level and derived from ZR singlets. Note that there is also the counterpart transition with change of parity seen at $q$ = 0 with optical measurements by Uchida $et$ $al.$\cite{Uchida1991} In the O $K$ edge data, as already noticed, this feature is hardly seen at increasing doping so that we cannot go deeper on this topic. 
 
\section{Conclusion}
\label{conc}

The present paper has  two main conclusions, one impacting on the RIXS method and the other on the physics of superconducting cuprates. The methodological contribution deals with the comparison of RIXS from undoped and doped cuprates. Indeed  in doped cuprates the RIXS at the oxyegn $K$ edge gives the same type of sampling of magnetic excitations as in the undoped cuprates provided the excitation is at the absorption pre-peak typical of doped systems. This implies that the comparison between doped and undoped systems must be done at different excitation energies. With this provision RIXS explores the bimagnon upper energy region around typically 450 meV and momentum within roughly one third of the BZ around the center. The dispersion, if any, is very small and the bimagnon intensity is almost constant with $\sigma$ polarization on going towards grazing incidence. The conclusion interesting the high T$_c$ cuprates comes from the exploitation of this method to study LSCO family from 0.1 to 0.22 doping. The main result is the almost constant weight of bimagnon and of higher even order spin excitations including the overdoped regime. This is accompanied by a damping of these excitations at higher dopings. These results contribute to a scenario in which damped spin excitations are present in the whole region of parameter space where superconductivity is found.

\acknowledgments
The work took enormous advantage from numerous stimulating and productive discussions with many colleagues and in particular with L. P. J. Ament, J. van den Brink, B. B\"uchner, J. Geck , J. Mesot. Moreover we are grateful to J. Mesot for having allowed us to use the bulk samples of LSCO. V.\,B. acknowledges the financial support from Deutscher Akademischer Austausch Dienst.


\begin{thebibliography}{10}%
\makeatletter
\providecommand \@ifxundefined [1]{%
 \ifx #1\undefined \expandafter \@firstoftwo
 \else \expandafter \@secondoftwo
\fi
}%
\providecommand \@ifnum [1]{%
 \ifnum #1\expandafter \@firstoftwo
 \else \expandafter \@secondoftwo
\fi
}%
\providecommand \enquote [1]{``#1''}%
\providecommand \bibnamefont  [1]{#1}%
\providecommand \bibfnamefont [1]{#1}%
\providecommand \citenamefont [1]{#1}%
\providecommand\href[0]{\@sanitize\@href}%
\providecommand\@href[1]{\endgroup\@@startlink{#1}\endgroup\@@href}%
\providecommand\@@href[1]{#1\@@endlink}%
\providecommand \@sanitize [0]{\begingroup\catcode`\&12\catcode`\#12\relax}%
\@ifxundefined \pdfoutput {\@firstoftwo}{%
 \@ifnum{\z@=\pdfoutput}{\@firstoftwo}{\@secondoftwo}%
}{%
 \providecommand\@@startlink[1]{\leavevmode}%
 \providecommand\@@endlink[0]{}%
}{%
 \providecommand\@@startlink[1]{%
  \leavevmode
  \pdfstartlink
   attr{/Border[0 0 1 ]/H/I/C[0 1 1]}%
   user{/Subtype/Link/A<</Type/Action/S/URI/URI(#1)>>}%
  \relax
 }%
 \providecommand\@@endlink[0]{\pdfendlink}%
}%
\providecommand \url  [0]{\begingroup\@sanitize \@url }%
\providecommand \@url [1]{\endgroup\@href {#1}{\urlprefix}}%
\providecommand \urlprefix [0]{URL }%
\providecommand \Eprint[0]{\href }%
\@ifxundefined \urlstyle {%
  \providecommand \doi [1]{doi:\discretionary{}{}{}#1}%
}{%
  \providecommand \doi [0]{doi:\discretionary{}{}{}\begingroup
  \urlstyle{rm}\Url }%
}%
\providecommand \doibase [0]{http://dx.doi.org/}%
\providecommand \Doi[1]{\href{\doibase#1}}%
\providecommand \bibAnnote [3]{%
  \BibitemShut{#1}%
  \begin{quotation}\noindent
    \textsc{Key:}\ #2\\\textsc{Annotation:}\ #3%
  \end{quotation}%
}%
\providecommand \bibAnnoteFile [2]{%
  \IfFileExists{#2}{\bibAnnote {#1} {#2} {\input{#2}}}{}%
}%
\providecommand \typeout [0]{\immediate \write \m@ne }%
\providecommand \selectlanguage [0]{\@gobble}%
\providecommand \bibinfo [0]{\@secondoftwo}%
\providecommand \bibfield [0]{\@secondoftwo}%
\providecommand \translation [1]{[#1]}%
\providecommand \BibitemOpen[0]{}%
\providecommand \bibitemStop [0]{}%
\providecommand \bibitemNoStop [0]{.\EOS\space}%
\providecommand \EOS [0]{\spacefactor3000\relax}%
\providecommand \BibitemShut [1]{\csname bibitem#1\endcsname}%
\bibitem{Tacon2011}%
  \BibitemOpen
  \bibfield{author}{%
  \bibinfo {author} {\bibfnamefont{M.}~\bibnamefont{Le~Tacon}}, \bibinfo
  {author} {\bibfnamefont{G.}~\bibnamefont{Ghiringhelli}}, \bibinfo {author}
  {\bibfnamefont{J.}~\bibnamefont{Chaloupka}}, \bibinfo {author}
  {\bibfnamefont{M.}~\bibnamefont{{Moretti Sala}}}, \bibinfo {author}
  {\bibfnamefont{V.}~\bibnamefont{Hinkov}}, \bibinfo {author}
  {\bibfnamefont{M.~W.}\ \bibnamefont{Haverkort}}, \bibinfo {author}
  {\bibfnamefont{M.}~\bibnamefont{Minola}}, \bibinfo {author}
  {\bibfnamefont{M.}~\bibnamefont{Bakr}}, \bibinfo {author}
  {\bibfnamefont{K.~J.}\ \bibnamefont{Zhou}}, \bibinfo {author}
  {\bibfnamefont{S.}~\bibnamefont{{Blanco-Canosa}}}, \bibinfo {author}
  {\bibfnamefont{C.}~\bibnamefont{Monney}}, \bibinfo {author}
  {\bibfnamefont{Y.~T.}\ \bibnamefont{Song}}, \bibinfo {author}
  {\bibfnamefont{G.~L.}\ \bibnamefont{Sun}}, \bibinfo {author}
  {\bibfnamefont{C.~T.}\ \bibnamefont{Lin}}, \bibinfo {author}
  {\bibfnamefont{G.~M.}\ \bibnamefont{{De Luca}}}, \bibinfo {author}
  {\bibfnamefont{M.}~\bibnamefont{Salluzzo}}, \bibinfo {author}
  {\bibfnamefont{G.}~\bibnamefont{Khaliullin}}, \bibinfo {author}
  {\bibfnamefont{T.}~\bibnamefont{Schmitt}}, \bibinfo {author}
  {\bibfnamefont{L.}~\bibnamefont{Braicovich}},\ and\ \bibinfo {author}
  {\bibfnamefont{B.}~\bibnamefont{Keimer}},\ }%
  \bibfield{journal}{%
  \Doi{10.1038/nphys2041}{\bibinfo {journal} {Nature Physics}}\ }%
  \textbf{\bibinfo {volume} {7}},\ \bibinfo {pages} {725} (\bibinfo {year}
  {2011})%
  \bibAnnoteFile{NoStop}{Tacon2011}%
\bibitem{Chen1991}%
  \BibitemOpen
  \bibfield{author}{%
  \bibinfo {author} {\bibfnamefont{C.~T.}\ \bibnamefont{Chen}}, \bibinfo
  {author} {\bibfnamefont{F.}~\bibnamefont{Sette}}, \bibinfo {author}
  {\bibfnamefont{Y.}~\bibnamefont{Ma}}, \bibinfo {author}
  {\bibfnamefont{M.~S.}\ \bibnamefont{Hybertsen}}, \bibinfo {author}
  {\bibfnamefont{E.~B.}\ \bibnamefont{Stechel}}, \bibinfo {author}
  {\bibfnamefont{W.~M.~C.}\ \bibnamefont{Foulkes}}, \bibinfo {author}
  {\bibfnamefont{M.}~\bibnamefont{Schluter}}, \bibinfo {author}
  {\bibfnamefont{S.-W.}\ \bibnamefont{Cheong}}, \bibinfo {author}
  {\bibfnamefont{A.~S.}\ \bibnamefont{Cooper}}, \bibinfo {author}
  {\bibfnamefont{L.~W.}\ \bibnamefont{Rupp}}, \bibinfo {author}
  {\bibfnamefont{B.}~\bibnamefont{Batlogg}}, \bibinfo {author}
  {\bibfnamefont{Y.~L.}\ \bibnamefont{Soo}}, \bibinfo {author}
  {\bibfnamefont{Z.~H.}\ \bibnamefont{Ming}}, \bibinfo {author}
  {\bibfnamefont{A.}~\bibnamefont{Krol}},\ and\ \bibinfo {author}
  {\bibfnamefont{Y.~H.}\ \bibnamefont{Kao}},\ }%
  \bibfield{journal}{%
  \Doi{10.1103/PhysRevLett.66.104}{\bibinfo {journal} {Physical Review
  Letters}}\ }%
  \textbf{\bibinfo {volume} {66}},\ \bibinfo {pages} {104} (\bibinfo {year}
  {1991})%
  \bibAnnoteFile{NoStop}{Chen1991}%
\bibitem{Chen1992}%
  \BibitemOpen
  \bibfield{author}{%
  \bibinfo {author} {\bibfnamefont{C.~T.}\ \bibnamefont{Chen}}, \bibinfo
  {author} {\bibfnamefont{L.~H.}\ \bibnamefont{Tjeng}}, \bibinfo {author}
  {\bibfnamefont{J.}~\bibnamefont{Kwo}}, \bibinfo {author}
  {\bibfnamefont{H.~L.}\ \bibnamefont{Kao}}, \bibinfo {author}
  {\bibfnamefont{P.}~\bibnamefont{Rudolf}}, \bibinfo {author}
  {\bibfnamefont{F.}~\bibnamefont{Sette}},\ and\ \bibinfo {author}
  {\bibfnamefont{R.~M.}\ \bibnamefont{Fleming}},\ }%
  \bibfield{journal}{%
  \Doi{10.1103/PhysRevLett.68.2543}{\bibinfo {journal} {Physical Review
  Letters}}\ }%
  \textbf{\bibinfo {volume} {68}},\ \bibinfo {pages} {2543} (\bibinfo {year}
  {1992})%
  \bibAnnoteFile{NoStop}{Chen1992}%
\bibitem{Peets2009}%
  \BibitemOpen
  \bibfield{author}{%
  \bibinfo {author} {\bibfnamefont{D.~C.}\ \bibnamefont{Peets}}, \bibinfo
  {author} {\bibfnamefont{D.~G.}\ \bibnamefont{Hawthorn}}, \bibinfo {author}
  {\bibfnamefont{K.~M.}\ \bibnamefont{Shen}}, \bibinfo {author}
  {\bibfnamefont{Y.-J.}\ \bibnamefont{Kim}}, \bibinfo {author}
  {\bibfnamefont{D.~S.}\ \bibnamefont{Ellis}}, \bibinfo {author}
  {\bibfnamefont{H.}~\bibnamefont{Zhang}}, \bibinfo {author}
  {\bibfnamefont{S.}~\bibnamefont{Komiya}}, \bibinfo {author}
  {\bibfnamefont{Y.}~\bibnamefont{Ando}}, \bibinfo {author}
  {\bibfnamefont{G.~A.}\ \bibnamefont{Sawatzky}}, \bibinfo {author}
  {\bibfnamefont{R.}~\bibnamefont{Liang}}, \bibinfo {author}
  {\bibfnamefont{D.~A.}\ \bibnamefont{Bonn}},\ and\ \bibinfo {author}
  {\bibfnamefont{W.~N.}\ \bibnamefont{Hardy}},\ }%
  \bibfield{journal}{%
  \Doi{10.1103/PhysRevLett.103.087402}{\bibinfo {journal} {Physical Review
  Letters}}\ }%
  \textbf{\bibinfo {volume} {103}},\ \bibinfo {pages} {087402} (\bibinfo {year}
  {2009})%
  \bibAnnoteFile{NoStop}{Peets2009}%
\bibitem{axes}%
  \BibitemOpen
  \bibfield{author}{%
  \bibinfo {author} {\bibfnamefont{C.}~\bibnamefont{Dallera}}, \bibinfo
  {author} {\bibfnamefont{E.}~\bibnamefont{Puppin}}, \bibinfo {author}
  {\bibfnamefont{G.}~\bibnamefont{Trezzi}}, \bibinfo {author}
  {\bibfnamefont{N.}~\bibnamefont{Incorvaia}}, \bibinfo {author}
  {\bibfnamefont{A.}~\bibnamefont{Fasana}}, \bibinfo {author}
  {\bibfnamefont{L.}~\bibnamefont{Braicovich}}, \bibinfo {author}
  {\bibfnamefont{N.~B.}\ \bibnamefont{Brookes}},\ and\ \bibinfo {author}
  {\bibfnamefont{J.~B.}\ \bibnamefont{Goedkoop}},\ }%
  \bibfield{journal}{%
  \bibinfo {journal} {Journal of Synchrotron Radiation}\ }%
  \textbf{\bibinfo {volume} {3}},\ \bibinfo {pages} {231} (\bibinfo {year}
  {1996})%
  \bibAnnoteFile{NoStop}{axes}%
\bibitem{polifemo}%
  \BibitemOpen
  \bibfield{author}{%
  \bibinfo {author} {\bibfnamefont{G.}~\bibnamefont{Ghiringhelli}}, \bibinfo
  {author} {\bibfnamefont{A.}~\bibnamefont{Tagliaferri}}, \bibinfo {author}
  {\bibfnamefont{L.}~\bibnamefont{Braicovich}},\ and\ \bibinfo {author}
  {\bibfnamefont{N.~B.}\ \bibnamefont{Brookes}},\ }%
  \bibfield{journal}{%
  \bibinfo {journal} {Review of Scientific Instruments}\ }%
  \textbf{\bibinfo {volume} {69}},\ \bibinfo {pages} {1610} (\bibinfo {year}
  {1998})%
  \bibAnnoteFile{NoStop}{polifemo}%
\bibitem{Chang2008}%
  \BibitemOpen
  \bibfield{author}{%
  \bibinfo {author} {\bibfnamefont{J.}~\bibnamefont{Chang}}, \bibinfo {author}
  {\bibfnamefont{C.}~\bibnamefont{Niedermayer}}, \bibinfo {author}
  {\bibfnamefont{R.}~\bibnamefont{Gilardi}}, \bibinfo {author}
  {\bibfnamefont{N.~B.}\ \bibnamefont{Christensen}}, \bibinfo {author}
  {\bibfnamefont{H.~M.}\ \bibnamefont{R\o{}nnow}}, \bibinfo {author}
  {\bibfnamefont{D.~F.}\ \bibnamefont{McMorrow}}, \bibinfo {author}
  {\bibfnamefont{M.}~\bibnamefont{Ay}}, \bibinfo {author}
  {\bibfnamefont{J.}~\bibnamefont{Stahn}}, \bibinfo {author}
  {\bibfnamefont{O.}~\bibnamefont{Sobolev}}, \bibinfo {author}
  {\bibfnamefont{A.}~\bibnamefont{Hiess}}, \bibinfo {author}
  {\bibfnamefont{S.}~\bibnamefont{Pailhes}}, \bibinfo {author}
  {\bibfnamefont{C.}~\bibnamefont{Baines}}, \bibinfo {author}
  {\bibfnamefont{N.}~\bibnamefont{Momono}}, \bibinfo {author}
  {\bibfnamefont{M.}~\bibnamefont{Oda}}, \bibinfo {author}
  {\bibfnamefont{M.}~\bibnamefont{Ido}},\ and\ \bibinfo {author}
  {\bibfnamefont{J.}~\bibnamefont{Mesot}},\ }%
  \bibfield{journal}{%
  \Doi{10.1103/PhysRevB.78.104525}{\bibinfo {journal} {Physical Review B}}\ }%
  \textbf{\bibinfo {volume} {78}},\ \bibinfo {pages} {104525} (\bibinfo {year}
  {2008})%
  \bibAnnoteFile{NoStop}{Chang2008}%
\bibitem{Okada2002}%
  \BibitemOpen
  \bibfield{author}{%
  \bibinfo {author} {\bibfnamefont{K.}~\bibnamefont{Okada}}\ and\ \bibinfo
  {author} {\bibfnamefont{A.}~\bibnamefont{Kotani}},\ }%
  \bibfield{journal}{%
  \Doi{10.1103/PhysRevB.65.144530}{\bibinfo {journal} {Physical Review B}}\ }%
  \textbf{\bibinfo {volume} {65}},\ \bibinfo {pages} {144530} (\bibinfo {year}
  {2002})%
  \bibAnnoteFile{NoStop}{Okada2002}%
\bibitem{Ellis2008}%
  \BibitemOpen
  \bibfield{author}{%
  \bibinfo {author} {\bibfnamefont{D.~S.}\ \bibnamefont{Ellis}}, \bibinfo
  {author} {\bibfnamefont{J.~P.}\ \bibnamefont{Hill}}, \bibinfo {author}
  {\bibfnamefont{S.}~\bibnamefont{Wakimoto}}, \bibinfo {author}
  {\bibfnamefont{R.~J.}\ \bibnamefont{Birgeneau}}, \bibinfo {author}
  {\bibfnamefont{D.}~\bibnamefont{Casa}}, \bibinfo {author}
  {\bibfnamefont{T.}~\bibnamefont{Gog}},\ and\ \bibinfo {author}
  {\bibfnamefont{Y.-J.}\ \bibnamefont{Kim}},\ }%
  \bibfield{journal}{%
  \Doi{10.1103/PhysRevB.77.060501}{\bibinfo {journal} {Physical Review B}}\ }%
  \textbf{\bibinfo {volume} {77}},\ \bibinfo {pages} {060501} (\bibinfo {year}
  {2008})%
  \bibAnnoteFile{NoStop}{Ellis2008}%
\bibitem{Ellis2010}%
  \BibitemOpen
  \bibfield{author}{%
  \bibinfo {author} {\bibfnamefont{D.~S.}\ \bibnamefont{Ellis}}, \bibinfo
  {author} {\bibfnamefont{J.}~\bibnamefont{Kim}}, \bibinfo {author}
  {\bibfnamefont{J.~P.}\ \bibnamefont{Hill}}, \bibinfo {author}
  {\bibfnamefont{S.}~\bibnamefont{Wakimoto}}, \bibinfo {author}
  {\bibfnamefont{R.~J.}\ \bibnamefont{Birgeneau}}, \bibinfo {author}
  {\bibfnamefont{Y.}~\bibnamefont{Shvyd'ko}}, \bibinfo {author}
  {\bibfnamefont{D.}~\bibnamefont{Casa}}, \bibinfo {author}
  {\bibfnamefont{T.}~\bibnamefont{Gog}}, \bibinfo {author}
  {\bibfnamefont{K.}~\bibnamefont{Ishii}}, \bibinfo {author}
  {\bibfnamefont{K.}~\bibnamefont{Ikeuchi}}, \bibinfo {author}
  {\bibfnamefont{A.}~\bibnamefont{Paramekanti}},\ and\ \bibinfo {author}
  {\bibfnamefont{Y.-J.}\ \bibnamefont{Kim}},\ }%
  \bibfield{journal}{%
  \Doi{10.1103/PhysRevB.81.085124}{\bibinfo {journal} {Physical Review B}}\ }%
  \textbf{\bibinfo {volume} {81}},\ \bibinfo {pages} {085124} (\bibinfo {year}
  {2010})%
  \bibAnnoteFile{NoStop}{Ellis2010}%
\bibitem{Braicovich2010}%
  \BibitemOpen
  \bibfield{author}{%
  \bibinfo {author} {\bibfnamefont{L.}~\bibnamefont{Braicovich}}, \bibinfo
  {author} {\bibfnamefont{J.}~\bibnamefont{van~den Brink}}, \bibinfo {author}
  {\bibfnamefont{V.}~\bibnamefont{Bisogni}}, \bibinfo {author}
  {\bibfnamefont{M.}~\bibnamefont{{Moretti Sala}}}, \bibinfo {author}
  {\bibfnamefont{L.~J.~P.}\ \bibnamefont{Ament}}, \bibinfo {author}
  {\bibfnamefont{N.~B.}\ \bibnamefont{Brookes}}, \bibinfo {author}
  {\bibfnamefont{G.~M.}\ \bibnamefont{{De Luca}}}, \bibinfo {author}
  {\bibfnamefont{M.}~\bibnamefont{Salluzzo}}, \bibinfo {author}
  {\bibfnamefont{T.}~\bibnamefont{Schmitt}},\ and\ \bibinfo {author}
  {\bibfnamefont{G.}~\bibnamefont{Ghiringhelli}},\ }%
  \bibfield{journal}{%
  \Doi{10.1103/PhysRevLett.104.077002}{\bibinfo {journal} {Physical Review
  Letters}}\ }%
  \textbf{\bibinfo {volume} {104}},\ \bibinfo {pages} {077002} (\bibinfo {year}
  {2010})%
  \bibAnnoteFile{NoStop}{Braicovich2010}%
\bibitem{LucioNote}%
  \BibitemOpen
  \bibinfo {note} {In collaboration with M. Le Tacon and B. Keimer, some of us (L.B. and G.G.) are working on paramagnons seen with RIXS at Cu $L_3$ in strongly overdoped systems.}%
  \bibAnnoteFile{Stop}{LucioNote}%
\bibitem{Ellis2011}%
  \BibitemOpen
  \bibfield{author}{%
  \bibinfo {author} {\bibfnamefont{D.~S.}\ \bibnamefont{Ellis}}, \bibinfo
  {author} {\bibfnamefont{J.}~\bibnamefont{Kim}}, \bibinfo {author}
  {\bibfnamefont{H.}~\bibnamefont{Zhang}}, \bibinfo {author}
  {\bibfnamefont{J.~P.}\ \bibnamefont{Hill}}, \bibinfo {author}
  {\bibfnamefont{G.}~\bibnamefont{Gu}}, \bibinfo {author}
  {\bibfnamefont{S.}~\bibnamefont{Komiya}}, \bibinfo {author}
  {\bibfnamefont{Y.}~\bibnamefont{Ando}}, \bibinfo {author}
  {\bibfnamefont{D.}~\bibnamefont{Casa}}, \bibinfo {author}
  {\bibfnamefont{T.}~\bibnamefont{Gog}},\ and\ \bibinfo {author}
  {\bibfnamefont{Y.-J.}\ \bibnamefont{Kim}},\ }%
  \bibfield{journal}{%
  \Doi{10.1103/PhysRevB.83.075120}{\bibinfo {journal} {Physical Review B}}\ }%
  \textbf{\bibinfo {volume} {83}},\ \bibinfo {pages} {075120} (\bibinfo {year}
  {2011})%
  \bibAnnoteFile{NoStop}{Ellis2011}%
\bibitem{Uchida1991}%
  \BibitemOpen
  \bibfield{author}{%
  \bibinfo {author} {\bibfnamefont{S.}~\bibnamefont{Uchida}}, \bibinfo {author}
  {\bibfnamefont{T.}~\bibnamefont{Ido}}, \bibinfo {author}
  {\bibfnamefont{H.}~\bibnamefont{Takagi}}, \bibinfo {author}
  {\bibfnamefont{T.}~\bibnamefont{Arima}}, \bibinfo {author}
  {\bibfnamefont{Y.}~\bibnamefont{Tokura}},\ and\ \bibinfo {author}
  {\bibfnamefont{S.}~\bibnamefont{Tajima}},\ }%
  \bibfield{journal}{%
  \Doi{10.1103/PhysRevB.43.7942}{\bibinfo {journal} {Physical Review B}}\ }%
  \textbf{\bibinfo {volume} {43}},\ \bibinfo {pages} {7942} (\bibinfo {year}
  {1991})%
  \bibAnnoteFile{NoStop}{Uchida1991}%
\end{thebibliography}
\end{document}